\documentclass{appolb}
\usepackage{epsfig}

\newcommand{\be}{\begin{equation}}
\newcommand{\ee}{\end{equation}}
\newcommand{\ba}{\begin{eqnarray}}
\newcommand{\ea}{\end{eqnarray}}
\newcommand{\ban}{\begin{eqnarray*}}
\newcommand{\ean}{\end{eqnarray*}}

\begin{document}

\title{Test of Bowler-Sinyukov Treatment \\ of Coulomb 
Interaction\thanks{Presented at the Fourth Workshop on Particle Correlations and Femtoscopy (WPCF2008), Cracow, Poland, September 11-14, 2008}}

\author{Rados\l aw Maj$^{\rm a}$, Stanis\l aw Mr\' owczy\' nski$^{\rm a,b}$ 
\address{$^{\rm a}$Institute of Physics, Jan Kochanowski University \\
ul.~\'Swi\c etokrzyska 15, PL - 25-406 Kielce, Poland \\
$^{\rm b}$Andrzej So\l tan Institute for Nuclear Studies \\
ul.~Ho\.za 69, PL - 00-681 Warsaw, Poland}}

\date{January 14, 2009}

\maketitle

\begin{abstract}

The Bowler-Sinyukov method to eliminate Coulomb interaction from
a two-particle correlation function is discussed and tested. 
\end{abstract}


\section{Introduction}

The correlation functions of two particles with `small' relative
momenta provide information about space-time characteristics of
particle's sources in high-energy nucleus-nucleus collisions 
\cite{Wiedemann:1999qn,Heinz:1999rw,Lisa:2005dd}. Within the 
standard `femtoscopy' method, one obtains parameters of a particle's 
source fitting experimental correlation functions with theoretical 
ones calculated in a given model. Since we usually deal with 
electrically charged particles, observed two-particle correlations 
are strongly influenced by the Coulomb interaction. The effect of 
the Coulomb force is eliminated from experimental data by means 
of the so-called Bowler-Sinyukov procedure 
\cite{Bowler:1991vx,Sinyukov:1998fc}. 

The femtoscopy was applied to a large volume of experimental data on
nucleus-nucleus collisions at SPS energy \cite{Heinz:1999rw}. The
spatial size of particle's sources appeared to be comparable to the
expected size a fireball created in nucleus-nucleus collisions while
the emission time of particles was significantly shorter. It was
predicted that at RHIC energies the emission time would be
significantly longer due to the long lasting hydrodynamic evolution
of the system created at the early stage of nucleus-nucleus
collisions \cite{Rischke:1995cm,Rischke:1996em}. To a big surprise
the experimental data obtained at RHIC \cite{Adler:2004rq,Adams:2004yc} 
show a very little, if any, change of the space-time characteristics of 
a fireball when compared to the SPS data. And in contradiction to
hydrodynamic models the emission time of particles appeared to be as 
short as 1 fm/c. Because of this surprising result, which is known as 
the `HBT Puzzle' \cite{Gyulassy:2001zv,Pratt:2003ij}, a reliability 
of the femtoscopy method was questioned. Very recently it has been 
shown that the hydrodynamic calculations can be modified to give 
rather short emission time of produced particles 
\cite{Broniowski:2008vp,Pratt:2008qv}, and thus the `HBT Puzzle' 
seems to be resolved. Nevertheless it is still of interest to 
quantitatively check the femtoscopy method. 

Our aim here is to test the Bowler-Sinyukov correction procedure 
which is used to eliminate the Coulomb interaction from the 
experimental data. The procedure assumes that the Coulomb effects 
can be factorized out. The correction's factor is calculated for 
a particle's source which is spherically symmetric and has zero 
lifetime. We examine the procedure applying it to the computed
Coulomb correlation functions of identical pions coming from 
anisotropic sources of finite lifetime. The effect of halo 
\cite{Nickerson:1997js} is also studied. We treat the computed 
Coulomb correlation functions as experimentalists deal with the 
measured correlation  functions. Thus, we extract the correlation 
function which is supposed to be free of the Coulomb interaction. 
However, in contrast to the situation of experimentalists we know 
actual parameters of particle sources which can be compared to the 
extracted ones. Our study is somewhat similar to that presented in 
\cite{Kisiel:2006is}.

We use the natural units, where $c = \hbar = 1$, and our metric
convention is $(+,-,-,-)$.


\section{Coulomb Correlation Function}
\label{sec-function}


We compute the correlation function using the well known Koonin formula
\cite{Koonin:1977fh}. Since the two particles of interest are described 
by means of nonrelativistic wave function, the computation is performed
in the center-of-mass frame of the pair, as the pair motion can be treated
as nonrelativistic in this frame. However, the source function, which
gives a probability to emit two particles at a given space-time distance
$(t,{\bf r})$, has to be transformed to the pair center-of-mass frame 
(where quantities are labeled with asterisks). The correlation function 
thus equals 
\be
\label{Koonin-relat-two}
C({\bf q}_*) = \int
d^3 r_* \: D_r({\bf r}_*) \:
|\varphi_{{\bf q}_*}({\bf r}_*)|^2 \;,
\ee
where $\varphi_{{\bf q}_*}({\bf r}_*)$ is the non-relativistic wave 
function of relative motion and $D_r({\bf r}_*)$ is the effective
source function 
\be
D_r({\bf r}_*) \equiv \int dt_* \, 
D_r(t_*,{\bf r}_*-{\bf v}_*t_*) \;.
\ee
where $D_r(t_*,{\bf r}_*)$ is  the `relative' source function. 
Obviously, the velocity of the pair in its center-of-mass frame 
vanishes (${\bf v}_*=0$). The `relative' source function is defined 
through the single-particle source function as
\be
\label{source-relat}
D_r(\mathbf{r},t) \equiv \int d^3R \, dT \:
D(\mathbf{R}-\frac{1}{2}\mathbf{r},T-\frac{1}{2}t) \:
D(\mathbf{R}+\frac{1}{2}\mathbf{r},T+\frac{1}{2}t) \;.
\ee
As a probability density, the source function is 
normalized to unity
\be
\label{norma}
\int d^3r \, dt \, D(t,\mathbf{r})= 
\int d^3r \, dt \, D_r(t,\mathbf{r}) =
\int d^3r \, D_r(\mathbf{r}) = 1 \;.
\ee

The Coulomb function of two non-identical particles interacting 
due to repulsive Coulomb force is well-known \cite{Schiff68} to
be
\be 
\label{coulomb-wave} 
\varphi_{\bf q}({\bf r})
= e^{- {\pi \eta \over 2 q}} \;
\Gamma (1 +i{\eta \over q} ) \; e^{i\bf qr} \;
F\big(-i{\eta \over q}, 1, i(qr - {\bf qr}) \big) \;,
\ee
where $q \equiv |{\bf q}|$ and $1/\eta$ is the Bohr radius of 
two-particle system which equals $\eta^{-1}_{\pi}=388$ fm for 
$\pi \pi$; $F$ denotes the hypergeometric confluent function. 
As we deal with pairs of identical bosons, the wave function 
$\varphi_{\bf q}({\bf r})$ is symmetrized. 

We choose the gaussian form of the single-particle source function
$D(t,{\bf r})$ but in order to easily transform it from the source
rest frame to the center-of-mass frame of the pair, we write it down 
in the Lorentz covariant form
\be
\label{source-cov}
D(x)=\frac{\sqrt{{\rm det}\Lambda}}{4\pi^2} \;
{\rm exp} [-\frac{1}{2}x_\mu \Lambda^{\mu\nu}x_\nu],
\ee
where $x^\mu=(t,{\bf r})$ is the position four-vector and 
$\Lambda^{\mu\nu}$ is the Lorentz tensor depending on 
the parameters $\tau$, $R_x$, $R_y$ and $R_z$, which 
characterize the life-time  and sizes of the source.
In the source rest frame the matrix is diagonal with the 
$\tau^{-2}$, $R_x^{-2}$, $R_y^{-2}$ and $R_z^{-2}$ along the
diagonal. The source function as written in  Eq.~(\ref{source-cov}) 
obeys the normalization condition (\ref{norma}) not only for the 
diagonal matrix $\Lambda$ but for non-diagonal as well. The source 
function (\ref{source-cov}) is evidently the Lorentz scalar that is
$$
D'(x') = \frac{\sqrt{{\rm det} \Lambda'}}{4 \pi^2}
\exp{[-\frac{1}{2} x'_\mu \Lambda'^{\mu\nu}x'_\nu]}
= \frac{\sqrt{{\rm det} \Lambda}}{4 \pi^2}
\exp{[-\frac{1}{2} x_\mu \Lambda^{\mu\nu}x_\nu]}
= D(x) \;,
$$
where $x'_\mu = L_{\mu}^{\;\;\nu}x_\nu$
and $\Lambda'^{\mu\nu} = L_{\;\;\sigma}^{\mu}\Lambda^{\sigma\rho}L_{\rho}^{\;\;\nu}$
with $L_{\sigma}^{\mu}$ being the matrix of Lorentz transformation. 
We note that ${\rm det} \Lambda' = {\rm det}L \: {\rm det} \Lambda
\: {\rm det}L^{-1} = {\rm det} \Lambda$. 

The correlation function of two identical noninteracting 
bosons should equal 2 at vanishing momentum 
($C_{\rm free} ({\bf q}=0)=2$) but free correlation functions 
extracted from experimentally measured ones appear to 
be significantly smaller than 2 at ${\bf q}=0$. There was 
introduced the idea of halo \cite{Nickerson:1997js} to explain 
this fact. It is assumed that only a fraction $f$ of particles 
contributing to the correlation function comes from the fireball 
while the remaining fraction ($1-f$) originates from long living 
resonances ($0\leq f \leq 1$). Then, we have two sources of 
the particles: the small one - the fireball and the big one 
corresponding to the long living resonances. The single-particle 
source function thus equals 
\be
\label{source-halo}
D(t,{\bf r}) = f \: D_f(t,{\bf r})+ (1-f)\: D_h(t,{\bf r}) \;,
\ee
where $D_f(t,{\bf r})$ and $D_h(t,{\bf r})$ represent the fireball 
and halo, respectively. If the halo radius $R_h$ is so large that 
$R_h^{-1}$ is below an experimental resolution of the relative 
momentum ${\bf q}$, the particles coming from halo do not 
contribute to the measured correlation function and one claims 
that $C_{\rm free}({\bf q} = 0) = 1 + \lambda$, where 
$\lambda \equiv f^2 < 1$. 

We have computed the Coulomb correlation functions for 
anisotropic gaussian sources of finite emission time. The halo 
has been also included. We use the Bertsch-Pratt coordinates 
\cite{Bertsch:1988db,Pratt:1986cc} 
{\it out, side, long}. These are the Cartesian coordinates, where 
the direction {\it long} is chosen along the beam axis ($z$), the 
{\it out} is parallel to the component of the pair momentum  
which is transverse to the beam. The last direction - {\it side} - 
is along the vector product of the {\it out} and {\it long} versors. 
So, the vector ${\bf q}$ is decomposed into the $q_o$, $q_s,$ and $q_l$ 
components. If the particle's velocity is chosen along the axis $x$, 
the out direction coincides with the direction $x$, the side direction 
with $y$ and the long direction with $z$. 


\section{The Bowler-Sinyukov procedure}
\label{sec-B-S}


The Coulomb effect is usually subtracted from the experimentally measured 
correlation functions by means of the Bowler-Sinyukov procedure. In the 
absence of halo, procedure assumes that the correlation function 
can be expressed as 
\be
\label{corr1}
C({\bf q})=K(q) \: C_{\rm free}({\bf q}) \;,
\ee
where $C_{\rm free}({\bf q})$ is the free correlation function
and $K(q)$ is the correction factor which depends only on 
$q \equiv |{\bf q}|$. The correction factor can be treated as
the Coulomb correlation function of two nonidentical particles 
of equal masses and charges. The function is, however, rather 
unphysical as the pair velocity with respect to the source is 
assumed to vanish even so the calculation is performed in the 
rest frame of the source where the source is assumed to be 
symmetric and of zero lifetime. The correction factor $K(q)$, 
which is described in detail in the Appendix to the paper 
\cite{Kisiel:2006is}, is computed as
\be
\label{B-S-pop}
K(q) = G(q) \int d^3r \:
D_r({\bf r}) \: |F(-\frac{i\eta}{q},1,i(qr-{\bf q}{\bf r}))|^2,
\ee
where $G(q)$ is the so-called Gamov factor equal
\be
\label{Gamov}
G(q) = {2 \pi \eta \over q} \,
{1 \over {\rm exp}\big({2 \pi \eta \over q}\big) - 1}
\ee
and $D_r({\bf r})$ describes the spherically symmetric gaussian 
source of zero lifetime and of the `effective' radius 
$R = \sqrt{(R_o^2 + R_s^2 + R_l)/3}$ where $R_o$, $R_s$ and 
$R_l$ are the actual source radii. 

To check the validity of Eq.~(\ref{corr1}), we have divided the 
computed Coulomb correlation function by the Correction factor 
$K(q)$. For the case of pion-pion correlations, the extracted 
free correlation function is almost identical with the actual 
correlation function of non-interaction particles. The procedure 
works very well even for strongly anisotropic sources. 

\begin{figure}[t]
\begin{minipage}{6cm}
\includegraphics*[width=6.4cm]{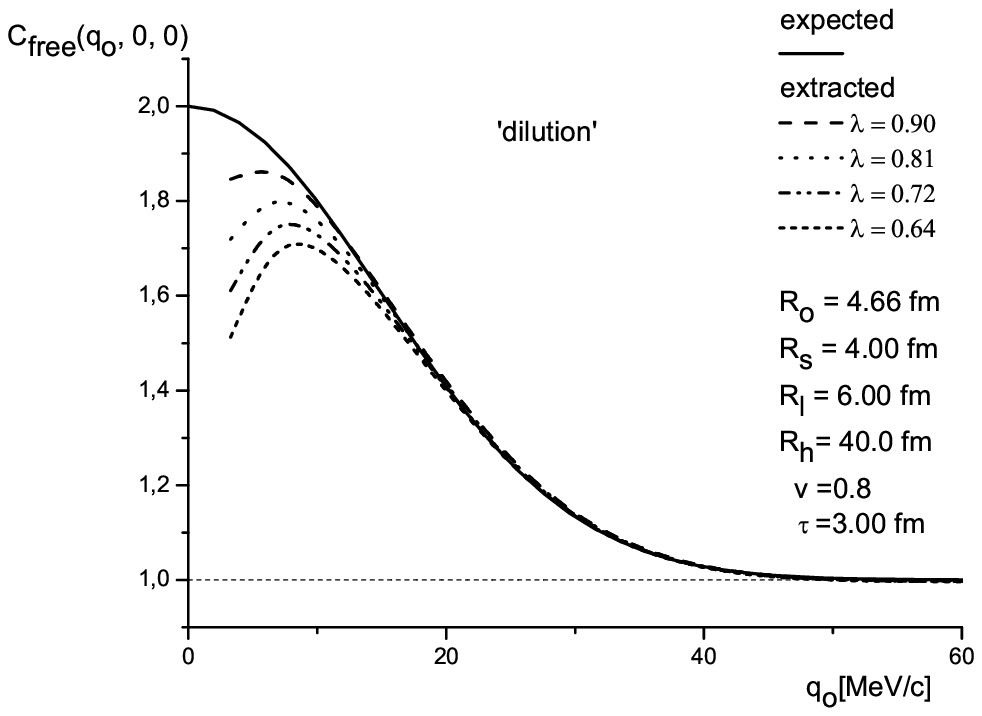}
\vspace{-5mm}
\end{minipage}
\hspace{1mm}
\begin{minipage}{6cm}
\includegraphics*[width=6.4cm]{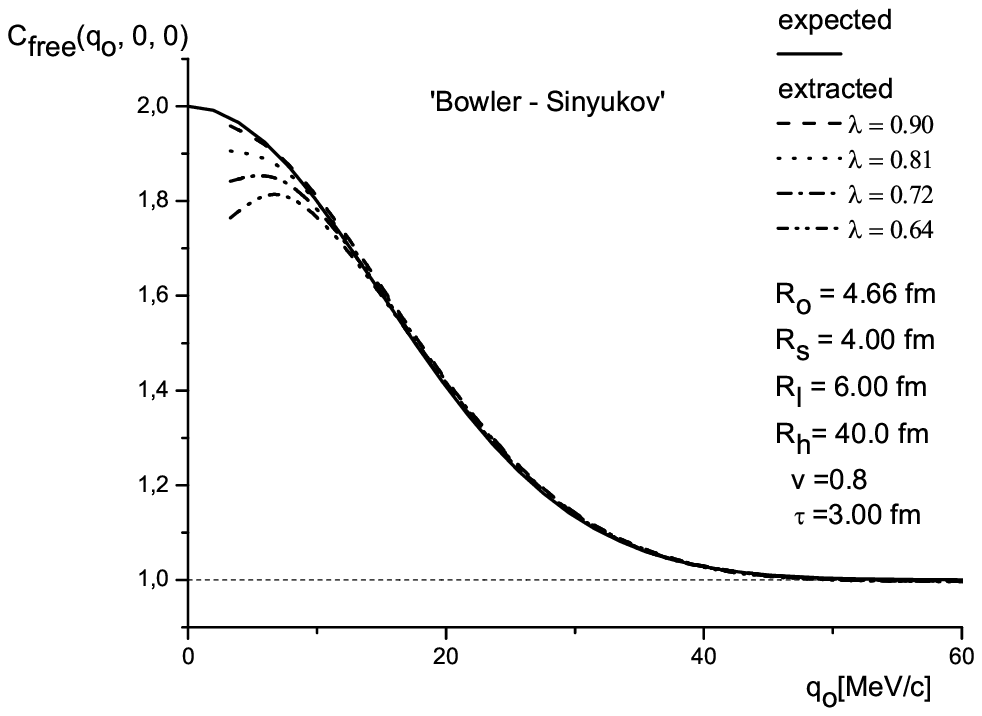}
\vspace{-5mm}
\end{minipage}
\caption{ The free correlation function $C_{\rm free} (q_o,0,0)$ 
extracted from the Coulomb correlation function by means of the 
dilution (left) and Bowler-Sinyukov (right) procedure for various 
$\lambda$. The expected free correlation function is also shown.}
\end{figure}

The situation is more complex when the halo is taken into account. 
We test two versions of the Bowler-Sinyukov procedure: the dilution 
method and the proper Bowler-Sinyukov one. The experimentally
measured correlation functions $C({\bf q})$ are fitted as
\begin{displaymath}
C({\bf q}) = \left\{
\begin{array}{cc}
\Big(1-\lambda +\lambda K(q) \Big)
\Big[1+ \lambda\big( C_{\rm free}({\bf q})-1\big)\Big]
& {\rm for \;\; dilution} \,, \\[3mm]
1-\lambda +\lambda K(q)C_{\rm free}({\bf q})
& {\rm for \;\; Bowler-Sinyukov} \,.
\end{array}
\right.
\end{displaymath}
The correlation function $C_{\rm free}({\bf q})$ extracted by means of the 
dilution and Bowler-Sinyukov procedures are shown in Fig.~1. The expected 
free function is shown for comparison. The source parameters are given 
in the figures. The parameter $\lambda$ is assumed to be known when 
$C({\bf q})$ are fitted. We show here only the function 
$C_{\rm free}({\bf q})$ for ${\bf q}=(q_o,0,0)$ which is crucial for the
emission time determination. As seen, the extracted correlation function 
is distorted at small relative momenta but the width of the correlation
function is unaltered and so are the source parameters. In our paper 
\cite{Maj-Mrow} we present a very detailed analysis of the Bowler-Sinyukov 
procedure. In particular, we show there that for the kaon-kaon correlations 
it works significantly worse than for the pion-pion ones.

We conclude our study as follows. In the absence of halo the Bowler-Sinyukov 
procedure works very well for $\pi \pi$ correlations. When the halo is 
taken into account the extracted correlation functions are distorted 
at small relative momenta but the source parameters are still reproduced 
accurately. 

\vspace{1mm}

We are very grateful to W.~Broniowski, W.~Florkowski, A.~Kisiel 
and R. Lednick\'y for numerous discussions. This work was partially 
supported by Polish Ministry of Science and Higher Education under 
grant N202 080 32/1843.


\end{document}